\documentclass[conference]{IEEEtran}
\IEEEoverridecommandlockouts
\usepackage{cite}
\usepackage{amsmath,amssymb,amsfonts}
\usepackage{algorithmic}
\usepackage{graphicx}
\usepackage{textcomp}
\usepackage{xcolor}
\usepackage{booktabs} 
\usepackage{caption} 
\usepackage{subcaption} 
\usepackage[hidelinks]{hyperref}

\def\BibTeX{{\rm B\kern-.05em{\sc i\kern-.025em b}\kern-.08em
    T\kern-.1667em\lower.7ex\hbox{E}\kern-.125emX}}

\usepackage{fancyhdr}
\makeatletter
\let\old@headrule\headrule
\renewcommand{\headrule}{\if@fancyplain\let\headrulewidth\plainheadrulewidth\fi\old@headrule}
\makeatother

\renewcommand{\headrulewidth}{0pt}

\makeatletter
\def\ps@IEEEtitlepagestyle{%
  \def\@oddhead{\mbox{}\scriptsize\rightmark \hfil}%
  \def\@evenhead{\scriptsize\thepage \hfil \leftmark\mbox{}}%
  \def\@oddfoot{\hfil \mbox{}\parbox{5.5in}{\centering
  \footnotesize \textcopyright 2024 IEEE. Personal use of this material is permitted.
  Permission from IEEE must be obtained for all other uses, in any current or future
  media, including reprinting/republishing this material for advertising or promotional
  purposes, creating new collective works, for resale or redistribution to servers or
  lists, or reuse of any copyrighted component of this work in other works.}\hfil \mbox{}}%
  \def\@evenfoot{\mbox{}\parbox{5.5in}{\centering}\hfil \mbox{}\hfil}%
}
\makeatother

\begin{document}

\title{It might be balanced, but is it actually good? \\ An Empirical Evaluation of Game Level Balancing\\

\thanks{This research was supported by the Volkswagen Foundation (Project: Consequences of Artificial Intelligence on Urban Societies, Grant 98555)\\ \\
979-8-3503-5067-8/   24/\$31.00~\copyright2024 IEEE \hfill}
}

\author{\IEEEauthorblockN{Florian Rupp}
\IEEEauthorblockA{\textit{Dep. of Computer Science} \\
\textit{Univ. of Appl. Sci. Mannheim}\\
Mannheim, Germany \\
f.rupp@hs-mannheim.de}
\and
\IEEEauthorblockN{Alessandro Puddu}
\IEEEauthorblockA{\textit{Institute for Applied AI} \\
\textit{Stuttgart Media University}\\
Stuttgart, Germany \\
ap110@hdm-stuttgart.de}
\and
\IEEEauthorblockN{Christian Becker-Asano}
\IEEEauthorblockA{\textit{Institute for Applied AI} \\
\textit{Stuttgart Media University}\\
Stuttgart, Germany \\
becker-asano@hdm-stuttgart.de}
\and
\IEEEauthorblockN{Kai Eckert}
\IEEEauthorblockA{\textit{Dep. of Computer Science} \\
\textit{Univ. of Appl. Sci. Mannheim}\\
Mannheim, Germany \\
k.eckert@hs-manneim.de}
}


\maketitle

\IEEEpubidadjcol

\begin{abstract}

Achieving optimal balance in games is essential to their success, yet reliant on extensive manual work and playtesting. To facilitate this process, the Procedural Content Generation via Reinforcement Learning (PCGRL) framework has recently been effectively used to improve the balance of existing game levels. 
This approach, however, only assesses balance heuristically, neglecting actual human perception.
For this reason, this work presents a survey to empirically evaluate the created content paired with human playtesting. Participants in four different scenarios are asked about their perception of changes made to the level both before and after balancing, and vice versa.
Based on descriptive and statistical analysis, our findings indicate that the PCGRL-based balancing positively influences players' perceived balance for most scenarios, albeit with differences in aspects of the balancing between scenarios.

\end{abstract}



\section{Introduction}
The balancing of a game greatly impacts the overall gaming experience. Unbalanced games lead to frustration and boredom and players will quit~\cite{becker_what_2020}.
Manually balancing a game, however, requires substantial work and testing in the development process of a game~\cite{schreiber_game_2021}. Automated balancing is therefore an important and active field of research~\cite{volz_demonstrating_2016,mesentier_silva_evolving_2019,karavolos_using_2018,pfau_dungeons_2020,jeon_raidenv_2023,rupp_geevo_2024}. Recently, we proposed a method using the Procedural Content Generation via Reinforcement Learning (PCGRL~\cite{khalifa_pcgrl_2020}) framework to balance existing tile-based game levels~\cite{rupp_balancing_2023},~\cite{rupp_simulation_2024}. To evaluate a level's balancing state, multiple simulations with heuristic agents are conducted to evaluate the balancing. A level is considered balanced when all agents win equally often.
The simulated balance, however, solely depends on the calculated heuristic. Therefore, we ask the question: Do humans perceive this calculated balance as intended?

To answer this question, this paper presents a survey based on human playtests of the generated content. Playing the game gives humans a more detailed insight in the actual balancing than just looking at the level. Each participant is asked to play one out of four randomly assigned scenarios, each consisting of an unbalanced and a balanced version of a level. After each playtest, participants are asked to answer questions about their perceptions of the balancing. 
As evaluation of game balance on an absolute scale leads to very subjective results~\cite{schreiber_game_2021}, we use a comparative rating~\cite{thurstone_law_1927} by asking participants how they rate the balance of a level in comparison to a different version of the level.
All tested levels along with the survey results are available on Github\footnote{https://github.com/FlorianRupp/pcgrl-balancing-empirical-evaluation}.
Our contributions are:
\begin{itemize}
    \item Design and conduction of a survey of automatically balanced game levels based on human playtests.
    \item A descriptive and statistical analysis of the survey data.
    \item A re-implementation of the game environment to make it accessible for human players.
\end{itemize}

\section{Related work}
Since the creation and balancing of game content requires a lot of manual effort including multiple human playtesters for instance, many works~\cite{volz_demonstrating_2016,mesentier_silva_evolving_2019,karavolos_using_2018,pfau_dungeons_2020,jeon_raidenv_2023,rogers_using_2023,rupp_geevo_2024} aim on automating this process using procedural content generation. 
Volz et al. introduce a method to algorithmically balance card games, however, they conclude: "[...] incorporating human perception of balancing is the only acceptable way [...]"~\cite[p.~276]{volz_demonstrating_2016}.

Similar to our work~\cite{rupp_balancing_2023,rupp_simulation_2024}, Karavolos et al.~\cite{karavolos_using_2018} use a PCG method for automated game level design using artificial agents to simulate game play. They conclude that this does not replicate human behavior one to one and must be evaluated accordingly.
In contrast, Pfau et al.~\cite{pfau_dungeons_2020} collect human game play data to model the behavior of humans to balance the content of a role-playing game.
Rogers et al.~\cite{rogers_using_2023} generate game economies and conclude, based on a survey, that the generated content can be controlled in terms of complexity and is perceived by humans in the intended way.
In their study participants were exposed to scenarios of varying complexity — either low or high — evaluated using Likert scales and Wilcoxon tests. In our study, we also compare two groups, however, we enable participants to experience both versions to enhance the accuracy of measuring the perceived differences.
Ashby et al. automatically generate quests for role playing games using large language models in combination with knowledge graphs~\cite{ashby_personalized_2023}. Participants in both an online survey and an additional user study were asked to evaluate the generated content by comparing it, among other things, with pre-existing human-created content.

\section{Background}
All levels we use in this work have been generated using our PCGRL-based method~\cite{rupp_balancing_2023,rupp_simulation_2024} for the same game setting within the Neural Massively Multiplayer Online (NMMO) environment~\cite{suarez_neural_2019}.

\subsection{Game Environment}
The game is a survival challenge on a 6x6 tile-based map where two players compete for food and water resources. Victory is achieved by either collecting five food resources or by 
surviving the opponent. There are four tile types: grass~\raisebox{-0.2em}{\includegraphics[width=1em]{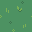}}, rock~\raisebox{-0.2em}{\includegraphics[width=1em]{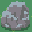}}, water~\raisebox{-0.2em}{\includegraphics[width=1em]{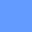}}, and food~\raisebox{-0.2em}{\includegraphics[width=1em]{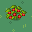}} (cf. Fig.~\ref{fig:scenarios}). Rock and water tiles impede movement.

Players make moves simultaneously, choosing from four directions: up, down, left, or right. Like in NMMO, their states include position, health, water, and food levels. Each turn, players lose water and food, with health depleting if both reach zero. When a player's health indicator reaches zero, the player has lost. Players can replenish food by consuming food tiles, which then become scrub tiles. There's a 2.5\% chance for scrub tiles to respawn as food. Water can be replenished by moving onto adjacent water tiles, which are never depleted. Health is restored gradually if food and water levels exceed 50\%.
In this work, the red player~\raisebox{-0.2em}{\includegraphics[width=1em]{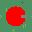}} is controlled by human players, while the yellow player~\raisebox{-0.2em}{\includegraphics[width=1em]{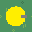}} is controlled by a heuristic agent. The heuristic implements a greedy strategy, always collecting the nearest available food resource.



\subsection{Level Generation and Balancing with PCGRL}
\label{sec:level-balancing}
The PCGRL framework~\cite{khalifa_pcgrl_2020} models level generation as a Markov decision process, leveraging reinforcement learning (RL). Within this iterative process, the RL agent receives rewards for modifying the current state until the level meets predefined constraints expressed through the reward function. Based on this approach, we can generate and balance levels independently. For balancing, only tile swaps are used as modifications. Rewards are computed by conducting multiple simulations, wherein heuristic agents are playing the game. The reward is determined by the frequency of each agent's victory in the game and can be understood as a static simulation-based evaluation function as defined in~\cite{yannakakis_experience-driven_2011}. This function assigns a value to each level that represents the balance in $[0,1]$, where 0.5 indicates equal win rates for both players, 0 and 1 indicate that a particular player wins each round.
Figure~\ref{fig:scenarios} shows the scenarios that we use in the survey. The first three scenarios use the balancer to balance a generated level, the last scenario uses a variation of the balancer to unbalance the balanced level from Scenario 2.

\section{Method}

\subsection{Survey Design}
\label{sec:survey}

\begin{figure}[htbp]
  \centering 
  \begin{subfigure}[b]{0.24\textwidth}
  \begin{subfigure}[b]{0.45\textwidth}\captionsetup[subfigure]{font=footnotesize}
    \includegraphics[width=\textwidth]{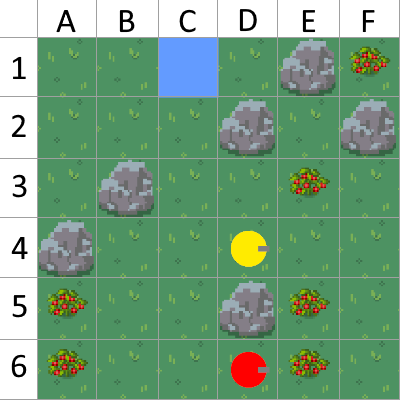}
    \centering \footnotesize Unbalanced, 0.3
  \end{subfigure}
  \hspace{0.3mm}
  \begin{subfigure}[b]{0.45\textwidth}
    \includegraphics[width=\textwidth]{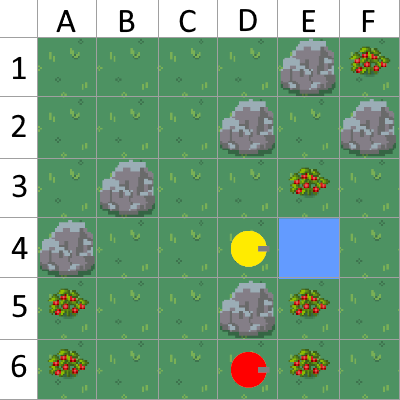}
    \centering \footnotesize Balanced, 0.5
  \end{subfigure}
  \caption{Scenario 1}
    \label{fig:s1}
  \end{subfigure}
\begin{subfigure}[b]{0.24\textwidth}
  \begin{subfigure}[b]{0.45\textwidth}
    \includegraphics[width=\textwidth]{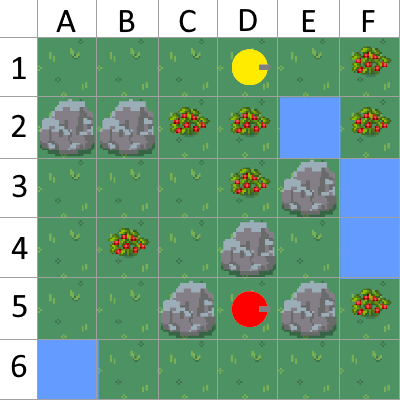}
    \centering \footnotesize Unbalanced, 0.0
  \end{subfigure}
    \hspace{0.3mm}
  \begin{subfigure}[b]{0.45\textwidth}
    \includegraphics[width=\textwidth]{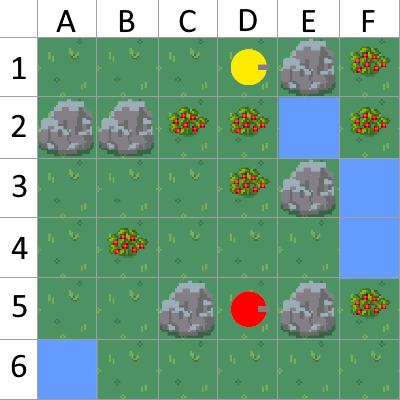}
    \centering \footnotesize Balanced, 0.5
  \end{subfigure}
  \caption{Scenario 2}
    \label{fig:s2}
  \end{subfigure}
  \begin{subfigure}[b]{0.24\textwidth}
  \vspace{2mm}
  \begin{subfigure}[b]{0.45\textwidth}
    \includegraphics[width=\textwidth]{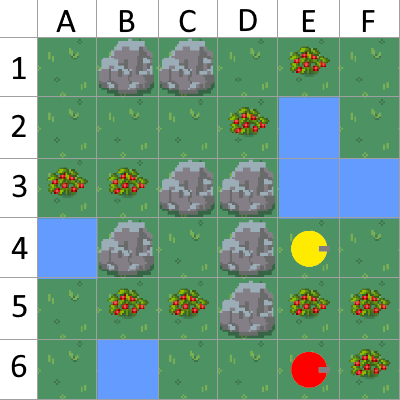}
    \centering \footnotesize Unbalanced, 0.2
  \end{subfigure}
    \hspace{0.3mm}
  \begin{subfigure}[b]{0.45\textwidth}
    \includegraphics[width=\textwidth]{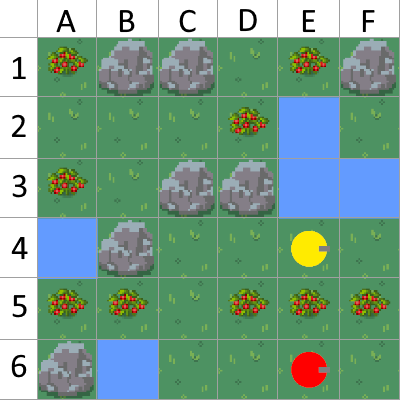}
    \centering \footnotesize Balanced, 0.5
  \end{subfigure}
  \caption{Scenario 3}
    \label{fig:s3}
  \end{subfigure}
  \hfill 
\begin{subfigure}[b]{0.24\textwidth}
  \begin{subfigure}[b]{0.45\textwidth}
    \includegraphics[width=\textwidth]{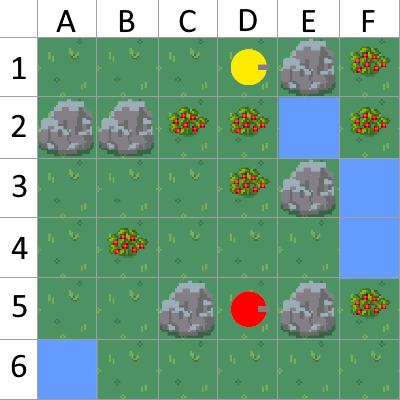}
    \centering \footnotesize Balanced, 0.5
  \end{subfigure}
    \hspace{0.3mm}
  \begin{subfigure}[b]{0.45\textwidth}
    \includegraphics[width=\textwidth]{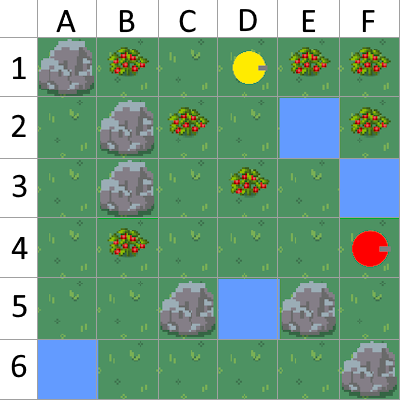}
    \centering \footnotesize Unbalanced, 0.0
  \end{subfigure}
  \caption{Scenario 4}
    \label{fig:s4}
  \end{subfigure}
  \caption{The four scenarios included for playtesting. Levels in scenarios 1-3 were balanced ($=0.5$) from a previously unbalanced version ($\neq 0.5$) by swapping tiles using the method from Rupp et al.~\cite{rupp_balancing_2023} (cf. Section~\ref{sec:level-balancing}). Scenario~4 presents the balanced version of scenario 2, which was subsequently unbalanced again using the PCGRL model.}
\label{fig:scenarios}
\end{figure}

\begin{figure}
\vspace{-6mm}
    \centering
    \includegraphics[width=0.8\linewidth]{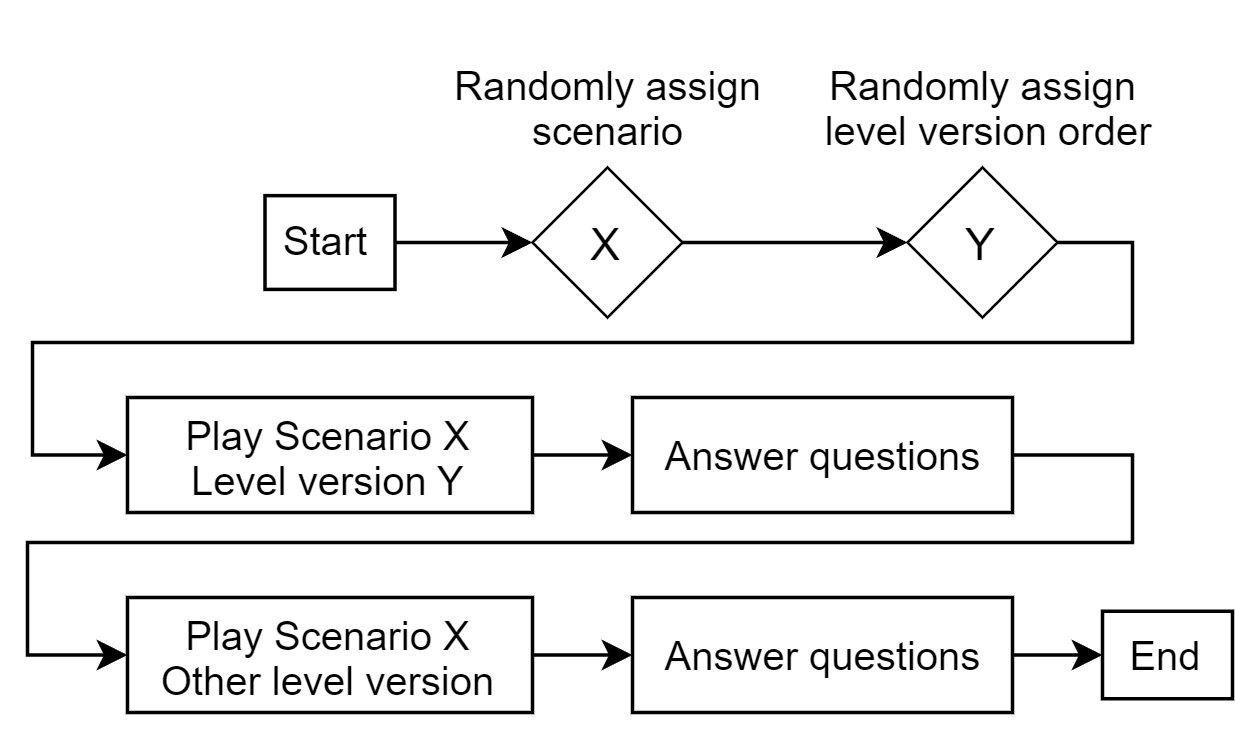}
  \caption{The design of the survey. Each participant is randomly assigned a scenario and which version of the level to play first.
  }
    \label{fig:survey}
\end{figure}

The survey's design plays a crucial role in empirically assessing balancing from the perspective of human players. Figure~\ref{fig:survey} outlines the survey process, which involves defining four scenarios, each consisting of paired unbalanced and balanced versions of the same level (Fig.~\ref{fig:scenarios}). Participants are randomly assigned a scenario at the onset.

To ensure participants' impartiality, the survey allows participants to sequentially experience both versions of a level without knowing which they are playing.
This is reinforced by randomizing the order in which participants encounter the versions, facilitating bi-directional evaluations.
Following each playtest, participants provide feedback on their perceptions of the balancing of the level they just experienced.

\begin{table*}
    \centering
    \caption{The descriptive and hypothesis testing results. For each item per scenario, the medians of the unbalanced (U) and the balanced (B) version are compared. The Wilcoxon signed-rank tests for each scenario compare the unbalanced with the balanced version per item. Significant values ($p \le 0.05$) and median changes are highlighted.}
    \begin{tabular}{lcccccccccccccccc} \toprule 
        & \multicolumn{4}{c}{Scenario 1} & \multicolumn{4}{c}{Scenario 2} & \multicolumn{4}{c}{Scenario 3} & \multicolumn{4}{c}{Scenario 4} \\

         \cmidrule(lr){2-5} \cmidrule(lr){6-9} \cmidrule(lr){10-13} \cmidrule(lr){14-17} 

        & \multicolumn{2}{c}{Median} & \multicolumn{2}{c}{Wilcoxon} & \multicolumn{2}{c}{Median} & \multicolumn{2}{c}{Wilcoxon} & \multicolumn{2}{c}{Median} & \multicolumn{2}{c}{Wilcoxon} & \multicolumn{2}{c}{Median} & \multicolumn{2}{c}{Wilcoxon}  \\
    
         \textbf{Item} &\textbf{U}& \textbf{B} & \textbf{W}&  \textbf{p} &\textbf{U}& \textbf{B} & \textbf{W}&\textbf{p} &\textbf{U}& \textbf{B} &  \textbf{W}&  \textbf{p}  &\textbf{B}& \textbf{U} &   \textbf{W}&\textbf{p} \\         
         
        \cmidrule(lr){1-1} \cmidrule(lr){2-3} \cmidrule(lr){4-5} \cmidrule(lr){6-7} \cmidrule(lr){8-9} \cmidrule(lr){10-11} \cmidrule(lr){12-13} \cmidrule(lr){14-15} \cmidrule(lr){16-17}
        
        Movement Difficulty Player &  1 & 1 & 35 & 0.225 &  \textbf{-2} & \textbf{0} & \textbf{5.5} & \textbf{0.00002} & 2 & 2 & \textbf{4.5}& \textbf{0.034}&  \textbf{-0.5} & \textbf{-1} & 15 & 0.357 \\
        Movement Difficulty Opponent & 1 & 1 & 20 & 0.740 & 2 & 2 & \textbf{0} & \textbf{0.007} &  \textbf{1} & \textbf{2} & \textbf{0}& \textbf{0.005}& \textbf{1} & \textbf{2} & 6.5 & 0.190 \\
        
        \cmidrule(lr){1-1} \cmidrule(lr){2-3} \cmidrule(lr){4-5} \cmidrule(lr){6-7} \cmidrule(lr){8-9} \cmidrule(lr){10-11} \cmidrule(lr){12-13} \cmidrule(lr){14-15} \cmidrule(lr){16-17}
        
        Amount Food Player & 0 & 0 & 3 & 0.180 &  \textbf{-1.5} & \textbf{1} & \textbf{7.5} & \textbf{0.0007} & 1 & 1 & 20 & 0.740 & -1 & -1  & 7 & 0.206 \\
        Amount Food Opponent & \textbf{0} & \textbf{-1} & \textbf{4} & \textbf{0.035} & 2 & 2 & 3 & 1.0 & 0 & 0 & 12 & 0.705 & 2 & 2 & 2 & 0.564 \\
        
        \cmidrule(lr){1-1} \cmidrule(lr){2-3} \cmidrule(lr){4-5} \cmidrule(lr){6-7} \cmidrule(lr){8-9} \cmidrule(lr){10-11} \cmidrule(lr){12-13} \cmidrule(lr){14-15} \cmidrule(lr){16-17}
        
        Amount Water Player & \textbf{-1.5} & \textbf{0} & \textbf{10.5} & \textbf{0.007} & 0 & 0 &  \textbf{0.0} & \textbf{0.006} & 0 & 0 & 22.5 & 1.0 & 0 & 0 & 9 & 0.194 \\
        Amount Water Opponent & \textbf{0.5} & \textbf{1} & 19.5 & 0.083 & \textbf{1} & \textbf{0} & \textbf{11} & \textbf{0.019} & 1 & 1 & 10.5 & 1.0 & 0 & 0 &  2 & 0.257 \\   
                 
        \cmidrule(lr){1-1}  \cmidrule(lr){2-5} \cmidrule(lr){6-9} \cmidrule(lr){10-13} \cmidrule(lr){14-17} \morecmidrules
        \cmidrule(lr){1-1}  \cmidrule(lr){2-5} \cmidrule(lr){6-9} \cmidrule(lr){10-13} \cmidrule(lr){14-17} 
        
        Critical Value for W & & & 29 & & &  & 52 & & &   & 73 & & & & 13 \\
        Samples & \multicolumn{4}{c}{16} & \multicolumn{4}{c}{20} & \multicolumn{4}{c}{23} & \multicolumn{4}{c}{12} \\
        \bottomrule
    \end{tabular}
    \label{tab:results}
\end{table*}

\subsubsection{Survey Questions}
Participants are asked to answer six questions (\emph{items}) for each level version, breaking down the abstract concept of balance into more tangible, game-related aspects for easier comprehension.
These questions assessed the proximity of the number of resources (food and water) to the player's spawn point and the difficulty of moving around the level due to impassable rocks and water tiles.
Additionally, participants evaluated their perception of their opponent's resource access based on its spawn position, resulting in six questions in total.
In the second level, participants answered the same six questions, but always in \emph{comparison} to the corresponding previous version, as comparative judgments are more intuitive for humans than absolute scales~\cite{thurstone_law_1927}. 


We use a five-point Likert scale for all questions.
Positioned in the middle of the scale is the choice indicating optimal balancing. To express either a surplus or a deficit, participants can choose between two levels respectively.
For example for the item \emph{Amount Food}, the scale ranges from \emph{way too few/none} (-2), \emph{too few} (-1), \emph{sufficient} (0), \emph{too much} (1), and \emph{way too much} (2).
The design is consistent across all questions in the first level, with slight variations in formulation based on the items. The mapping to numerical labels is not shown to participants.
The complete catalog of questions and possible choices is provided in the Github repository (see page 1). For instance, the corresponding question for the item \emph{Amount Food} is: \emph{How would you rate the amount of berry bush tiles (food) near your starting position?}


\subsection{Data Analysis}
\label{sec:analysis}

The data analysis is twofold: firstly, a descriptive comparison of median values per item before and after balancing, and secondly, Wilcoxon signed-rank hypothesis tests to assess the statistical significance of items before and after balancing.

\paragraph{Data preparation} To accurately compare median values and conduct hypothesis tests based on the previously mentioned relative data (cf.~\ref{sec:survey}), it is imperative to convert it into absolute representations.

Per item we create a proxy to obtain the absolute value for the second level played $L2$ using the absolute value from the first level $L1$ and the relative one from the second level $\Delta L2$. This can be expressed with the formula $f(L1, \Delta L2) = L1 + \Delta L2$. To revert the data for cases where the balanced version has been played first we retrieve the absolute value using $f(L2, \Delta L1)$. For instance, if a participant rates the available amount of food for the unbalanced version $L1$ as \emph{way too much} (2) and in the balanced version $\Delta L2$ as \emph{fewer} (-1) this results in a score of 1, being still (\emph{too much}), but not \emph{way too much} anymore.
We fix the scale to the interval of $[-2,2]$ and round values exceeding this interval to the closest possible value.

\paragraph{Hypothesis tests} 
To evaluate changes in participants' perceptions of items before and after balancing, we utilize the two-tailed Wilcoxon signed-rank test, suitable for paired, ordinal, and non-normally distributed data. A preliminary Shapiro-Wilk test revealed that neither case exhibited a normal distribution for both distributions respectively.
The null hypothesis $H_0$ states that the data distribution does not significantly change after balancing the level. We select a significance level of 0.05. If the test's p-value falls below 0.05 and the test statistic $W$ is less than the critical value $W$ determined by the sample size, we reject $H_0$ and accept $H_1$, indicating a change in the data distribution.
Despite multiple tests, we don't apply any measure to correct the p-values, such as the Bonferroni correction, as this would increase the risk of Type II errors due to the relatively small sample size in each scenario.



\section{Results, Discussion and Limitations}

Participants were primarily recruited from two groups: students and academic staff.
We only included complete responses for further analysis, for a total of 71 valid responses. The number of submissions per scenario varies as some surveys were not fully completed.
Table~\ref{tab:results} presents the descriptive results (median) per item in each scenario, along with the results of Wilcoxon signed-ranked tests comparing the data before and after balancing.

The results indicate significant changes in the perception of particular items across scenarios 1, 2, and 3. Notable findings underscore variability in the perception of items across different levels.
For instance, in scenario 1 (Fig.~\ref{fig:s1}), there is a significant increase in the player's accessibility to the water resource (\emph{Amount Water Player}). Also in the descriptive results is a noticeable improvement, as evidenced by the median shifting from -1.5 (\emph{too few$\,$/$\,$way to few}) to 0 (\emph{adequate}) in the balanced version. This suggests that participants perceived the single water tile swap as affecting the level's balance positively.
Conversely, in scenario 3 (Fig.~\ref{fig:s3}), perceptions of player movement difficulty varied, with no discernible shift in perceptions of water accessibility.

In scenario 2 (Fig.~\ref{fig:s2}), numerous items exhibited significant perceptual changes after balancing. Particularly noteworthy are the shifts in the items \emph{Movement Difficulty Player} and \emph{Amount Food Player}. Additionally, descriptive results suggest changes in the distribution towards a better balancing. For instance, the median value shifted for the item \emph{Movement Difficulty Player} from \emph{very inconvenient}~(-2) to \emph{adequate}~(0) indicating balance after the balancing procedure. A comparable improvement can be observed for the \emph{Amount Water Opponent} item, where the access to water resources has been reduced from \emph{slightly more}~(1) to \emph{same amount}~(0). The perception of \emph{Amount Food Player} also shows significant variation; however, the descriptive results suggest that the player now has slightly excessive food resources. Nevertheless, in sum this improvement remains positively valued.
In scenario 4, we tested the RL's capability to unbalance the initially balanced level from scenario 2 in favor of the opponent. Although players did not perceive statistically significant differences, a descriptive contrast emerged: the player's movement slightly worsened, while the opponent's movement slightly improved.

Not all balanced versions were universally perceived as improved towards the center of the Likert scale. In scenario 1 the perception of the \emph{Amount Food Opponent} is slightly worse in the balanced version.
Especially in scenario 3 regarding the item \emph{Movement Difficulty Opponent}, participants rated their opponent's freedom of movement even better than in the unbalanced version, as indicated by the median and the significance of the hypothesis test. Despite this, both player and opponent are now rated similarly. This suggests that while movement is still relatively easy, it is now comparably easy for both players.

A limitation arises from the time-consuming nature of playtests involving humans, restricting our ability to evaluate only a small subset of levels, despite the potential to generate thousands using the PCGRL method.
The balance across all levels was heuristically estimated through multiple simulations involving scripted agents. These agents consistently behave identically, varying only due to the probabilistic nature of the game environment.
However, humans adapt their strategies as they play and learn from each experience. Consequently, they adjust their strategy upon replaying a level. This contrasts with the setup when balancing the levels, where two static scripted agents of always precisely equal skill face each other.

For these reasons, we aim to encourage authors not only to rely on the validity of the generated content, but also to include human feedback in a method's evaluation process.


\section{Conclusions}
In this paper, we designed and conducted a survey paired with human playtesting, to empirically evaluate automated game balancing based on heuristics and reinforcement learning.
Participants were presented with one of four scenarios, each consisting of a pair of unbalanced and balanced level versions in random order.
Descriptive analysis coupled with hypothesis testing revealed significant differences in the perceived balance distribution of levels pre- and post-automated balancing. 
Notably, participants perceived balance differently across various aspects such as resource availability and freedom of movement within each scenario.
So, is the balancing actually good? Our findings suggest that our balancing approach influences balance perception in most cases positively; however, human perceptions may differ in certain aspects, depending on the level. We conclude that while the investigated automated method can balance levels accordingly to reduce the need for manual human work, a final human evaluation still remains essential.
Moreover, we believe that the survey design presented can be applied to the empirical evaluation of content that has undergone procedural optimization for a specific objective. In essence, it is applicable to scenarios where both the original and optimized versions of the content are available for comparison.

\bibliographystyle{unsrt}
\bibliography{library.bib}

\end{document}